\documentclass[letterpaper,12pt]{article}
\usepackage{hyperref}
\usepackage{graphicx}
\usepackage{amssymb}
\usepackage{floatflt}
\usepackage{hyperref}
\usepackage{multicol}
\begin{document}
\thispagestyle{empty}

\begin{center}
\Large{The ngVLA's Role in Exoplanet Science: Constraining Exo-Space Weather \\ 

\vspace*{0.5cm} 
\textit{\large A white paper submitted to the National Academy of Science Committee on Exoplanet Science Strategy } }
\normalsize

\centerline{March 9, 2018}

\vspace{1cm}

Rachel A. Osten\\
Space Telescope Science Institute \& Johns Hopkins University \\
(410) 338-4762 \\
Baltimore, MD \\
osten@stsci.edu\\
\vspace{0.5cm}
Michael K. Crosley \\
Johns Hopkins University \\
Baltimore, MD \\
\vspace{0.5cm}
Manuel G\"{u}del \\
University of Vienna \\
Vienna, Austria \\
\vspace{0.5cm}
Adam Kowalski \\
University of Colorado, Boulder \& National Solar Observatory \\
Boulder, CO \\
\vspace{0.5cm}
Joe Lazio \\
Jet Propulsion Laboratory \\
Pasadena, CA \\
\vspace{0.5cm}
Jeffrey Linsky \\
JILA, University of Colorado, Boulder \\
Boulder, CO \\
\vspace{0.5cm}
Eric Murphy \\
National Radio Astronomy Observatory \\
Charlottesville, VA \\
\vspace{0.5cm}
Stephen White \\
Air Force Research Lab \\
Albequerque, NM \\

\end{center}

\newpage

\setcounter{page}{1}




\section{ What Radio Observations Can Bring to Exoplanet Science Studies}


A star can influence its near environment through radiation and particles. Stellar flaring and associated coronal mass ejections, a steady stellar wind, and star-planet magnetospheric interactions are all important factors in the space weather environments of exoplanets. Chapter 6 of NASA's Astrobiology Strategy (\href{https://nai.nasa.gov/media/medialibrary/2015/10/NASA_Astrobiology_Strategy_2015_151008.pdf}{link)} specifically calls this out, asking the question \\[-5mm]
\begin{quotation}
What are the properties of the host star that are conducive to or prevent the formation of a habitable world: age, size, and elemental composition of the star, its activity, and the properties of nearby stars?
\end{quotation}
Radio observations provide direct observations on the particle environment created by the star. The science described in this white paper is needed to realize that fuller understanding of exoplanet ecosystems. 

An international community of astronomers is currently engaged in formulating next generation science questions motivated by the scientific progress made by the Jansky Very Large Array (JVLA) and the Atacama Large Millimeter Array (ALMA) radio telescopes. The vision is for a next generation VLA, optimized to perform imaging of thermal emission down to milliarcsecond scales, with sensitivities roughly ten times the current collecting area of the JVLA and ALMA, and operating in the frequency range 1 - 115 GHz. This array bridges the gap between ALMA, working primarily at shorter wavelengths, and the future SKA1, optimized for longer wavelengths.    More information on the project as a whole can be found at \url{http://ngvla.nrao.edu}. The project has been increasing in momentum in the last few years, with an ngVLA memo series (\url{http://ngvla.nrao.edu/page/memos}), capacity-limited community workshops, and funding for the community to develop science topics and engineers to advance antenna and array designs. With construction anticipated to commence in 2024, early science would start in 2028, with full array operations in 2034.       

The contents of this white paper are based substantially on work reported in Osten \& Crosley (2017) ngVLA Memo \#31, available at \url{http://library.nrao.edu/public/memos/ngvla/NGVLA_31.pdf}.    \\[-8mm]

\subsection{Mass Loss on the Lower Half of the Main Sequence}
Stellar winds affect the migration and/or evaporation of exoplanets (Lovelace et al. 2008), and are 
important not only for understanding stellar rotational evolution, but also for their influence
on planetary dynamos (Heyner et al. 2012).
A direct consequence of rotational spin-down is the decline of activity and therefore XUV radiation.
This evolution is highly non-unique in the early stellar evolution but has very
  important consequences for planetary thermal escape (e.g., Tu et al. 2015; plus
  chemical processing). Therefore stellar wind observations for stars with known
  rotation periods would provide important clues as to how much angular momentum
  is being transported away (presently, we need models including magnetic fields
  and ``Alfven surfaces" to link spin-down to winds, with some free parameters).
At present only indirect measures of cool stellar mass loss inform these topics.  
Mass loss 
in the cool half of the HR diagram, along/near the main sequence, has been notoriously difficult to detect directly, 
due in part to the 
much lower values of mass loss here compared to other stellar environments 
($\dot M$ of $2\times$ 10$^{-14}$ solar masses per year for the Sun). 

Cool stellar mass loss is characterized by an ionized stellar wind, whose radio flux can have a $\nu^{0.6}$ or
$\nu^{-0.1}$ dependence if in the optically thick or thin regime, respectively.  
A direct measurement of stellar 
mass loss through its radio signature would be a significant leap forward not only for understanding the plasma physics of the 
stars themselves, but also for understanding what kind of environment those stars create. 
Previous attempts at a direct detection of cool stellar mass loss via radio emission have led to upper limits typically
 three to four orders of magnitude higher than the Sun's present-day mass loss, while indirect methods find  evidence
for mass loss rates comparable to or slightly higher than the Sun's present day mass loss rate (up to $\sim$80 times solar
$\dot M$.)

Measurements with the ngVLA can provide vital information on  stellar mass loss as crucial input to exoplanet studies in two key areas:
\begin{itemize}
\item \textit{Increasing the Look-back Time for Studying the Young Sun} There is tension between the results returned from the astrospheric detection method and the mass loss expected from stellar rotational evolution models. The results of Johnstone et al. (2015) suggest that the Sun at young ages should have had a mass loss rate roughly an order of magnitude higher than it does today.
Fichtinger et al. (2017) presented recent results obtained from the JVLA and ALMA for a sample of nearby stars. Results are shown in Figure~1. Results from the ngVLA will be able to make a significant leap in understanding the mass loss history of the Sun
through its two order of magnitude improvement over what can be achieved at present with the JVLA, a combination of the increase in sensitivity, observing frequency, and integration time. In addition, non-detections provide quantitative constraints
on the mass loss rate, a marked difference from the astrospheric absorption method. 
\begin{figure}
\begin{minipage}[c]{0.6\linewidth}
\includegraphics[scale=0.4]{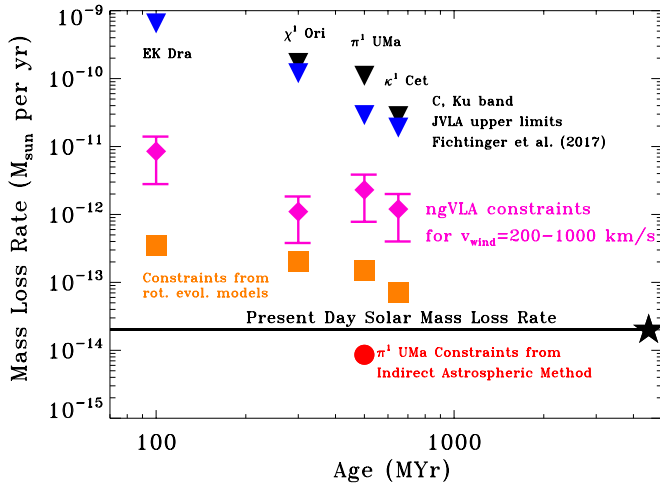}
\end{minipage}
\begin{minipage}[c]{0.39 \linewidth}
\caption{Current mass loss constraints for nearby solar analogs, along with prospects achievable with the ngVLA. Black and blue
triangles indicate upper limits from Fichtinger et al. 2017 for spherically symmetric winds; orange squares show mass loss constraints from the rotational 
evolution models of Johnstone et al. (2015); and red circle shows detection from the indirect method of inferring mass loss via astrospheric absorption (Wood et al. 2014).
Magenta diamonds and error bars display the grasp of the ngVLA, for wind velocities spanning 200-1000 km s$^{-1}$. 
}
\end{minipage}
\end{figure}
\item \textit{Mass Loss from M Dwarfs in the Solar Neighborhood} Consideration of the steady stellar wind as well as time-varying flares and coronal mass ejections from M dwarfs are vital components to understanding
the complex intertwining that may or may not turn a potentially habitable planet into an inhabited planet.The method of detecting radio emission from nearby M dwarfs will be an important contributor to understanding how the wind environment of
nearby M dwarfs contributes to the habitability of orbiting exoplanets. A high rate of coronal mass ejections may also act like a wind near planets, but may also be responsible for spin-down instead of a steady wind. Figure~2 shows the sensitivity of the ngVLA to detecting radio emission from an ionized stellar wind for M dwarfs within 10 pc, as well as the few current observational constraints on M dwarf mass loss. 
This method is most sensitive to the nearest stars, of which there are $\sim$270 M dwarfs within 10 pc.  The interpretation of the radio emission requires disentangling any wind-related signature from other magnetic activity signatures, making this technique potentially more optimistic for studies of low activity M dwarfs than higher activity stars. 
\begin{figure}
\begin{minipage}[c]{0.6\linewidth}
\hspace*{-1.5cm}
\includegraphics[scale=0.4]{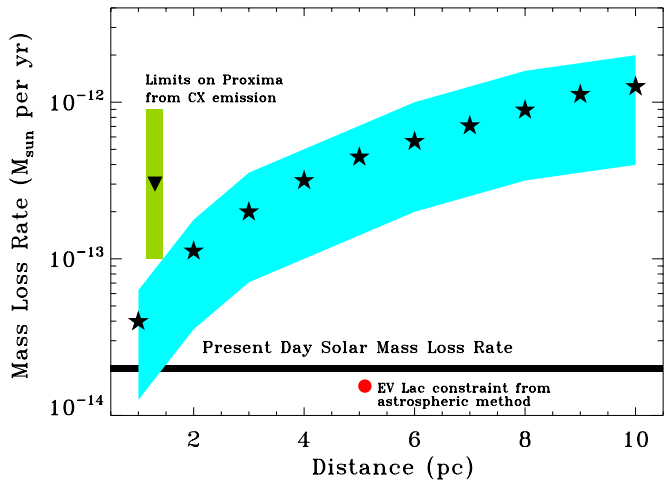}
\vspace*{-1.5cm}
\end{minipage}
\begin{minipage}[c]{0.39\linewidth}
\caption{Stellar mass loss rate constraints for M dwarfs in the solar neighborhood.  Stars and cyan shaded region gives area able to be constrained by 
ngVLA observations. Current observational limits on M dwarf wind mass loss are also indicated: the green triangle indicates the upper limit for Proxima based on limits
from charge exchange emission (Wargelin \& Drake 2001). The stellar wind mass loss rate implied by the detection of
Lyman $\alpha$ astrospheric absorption towards the M dwarf EV~Lac at 5.1 pc is also indicated.
}
\end{minipage}

\end{figure}

\end{itemize}

While measurements of radio emission from point sources are easy to undertake, their interpretation is complicated. The calculations above assume 
optically thick wind emission. However, gyrosynchrotron emission commonly seen in active stars has a peak frequency near 10 GHz, and flat or slightly negative spectrum at higher frequencies.  At even higher frequencies probed by ALMA, stellar chromospheric emission is detected.  Thus the region between 10 and 100 GHz is ideal for searching for wind emission from nearby stars, but careful interpretation to include the potential contribution of other emission processes is needed.\\[-8mm]


\subsection{Particles and Fields: Contribution to Exo-Space Weather}
%
Radio observations provide a unique way to characterize the nature
of the accelerated electron population near the star. This is
important for understanding the radiation and particle environment in which close-in exoplanets
are situated. As an example, Segura et al. (2010) investigated the impact on the ozone layer of a superflare from an M dwarf. The energetic particles can 
deplete the ozone layer of a planet in the habitable zone of an Earthlike planet around an M dwarf experiencing a superflare. 
Radio observations provide the clearest signature of accelerated particles and shocks in stars arising from transient magnetic reconnection, and provide more realistic constraints on these factors than scaling by solar values (Osten \& Wolk 2017). Optically thin gyrosynchrotron emission constrains the index of the accelerated particles, and 
also enables a constraint on magnetic field strengths in the radio-emitting source (Smith et al. 2005, Osten et al. 2016). The peak frequency is about 10 GHz, so observations
at higher frequencies are necessary to disentangle optical depth effects from the interpretation.

Radio constraints on accelerated particle characteristics are  the only 
 measurement technique that will yield results for planet-hosting stars in the solar neighborhood.
For radio wavelength observations,  the time-dependent response of radio emission 
reveals the changing nature of the magnetic field strength and number and distribution of accelerated particles.
This would
provide constrains on the accelerated particle population of close-in exoplanets unavailable from any other
observational method; such a constraint is necessary to perform detailed modeling of the atmospheres of such exoplanets,
due to the influence of accelerated particles in affecting the chemical reactions in terrestrial planet atmospheres 
(Jackman et al. 1990). There are numerous M dwarfs detectable with the ngVLA at the low radio luminosity end near
 10$^{11-12}$ erg s$^{-1}$ Hz$^{-1}$ for which an analysis similar to that performed in Smith et al. (2005) and Osten et al. (2016) could be performed. 
 The integrated radio flare energy provides a constraint on the accelerated particle
kinetic energy, under the assumption of optically thin emission. 
The radio light curve provides constraints on  both
the index of the distribution of accelerated particles, as well as the magnetic field strength in the radio-emitting source. 

\begin{figure}
\includegraphics[scale=0.35]{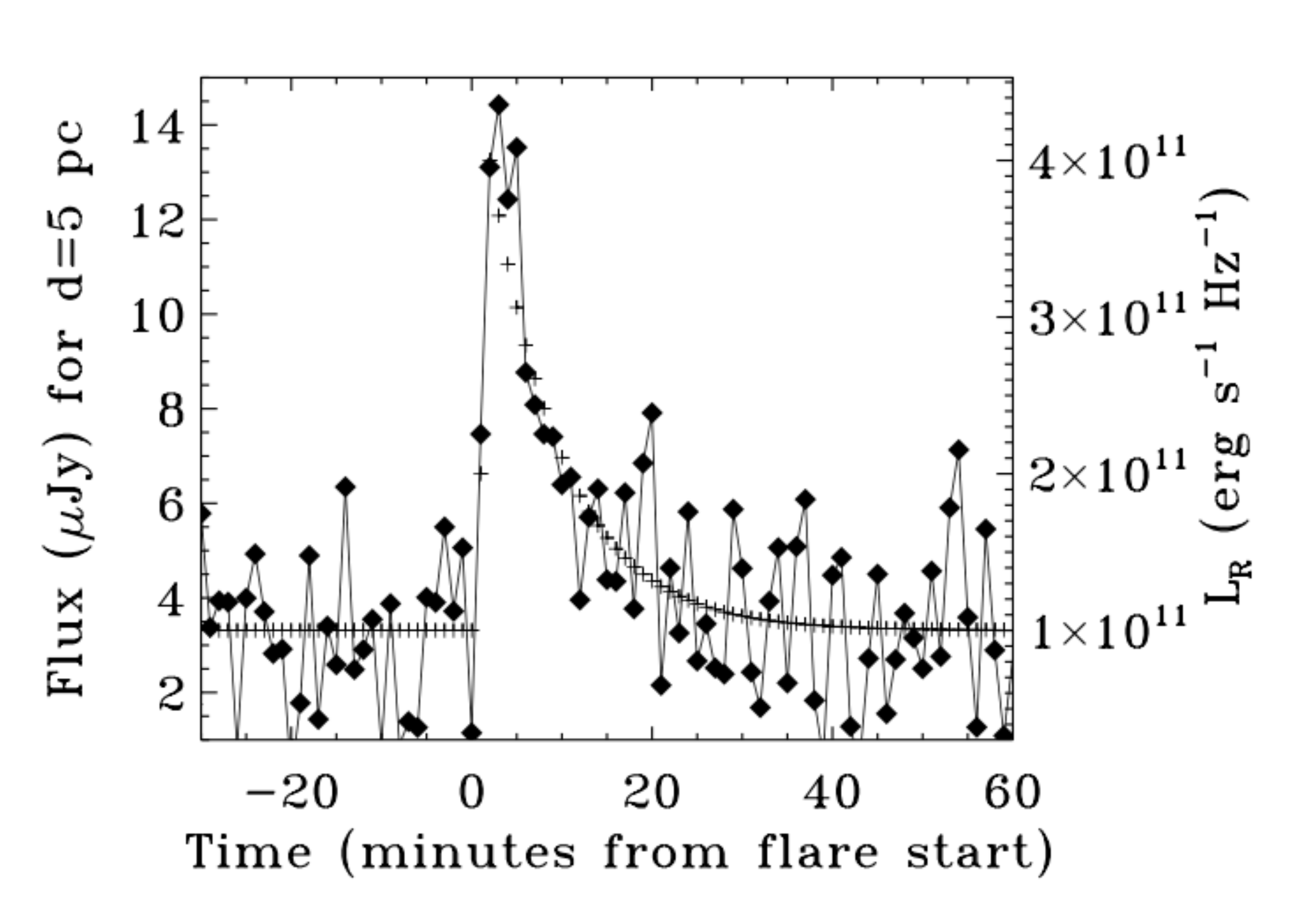}
\hspace*{-1.5cm}
\includegraphics[scale=0.3]{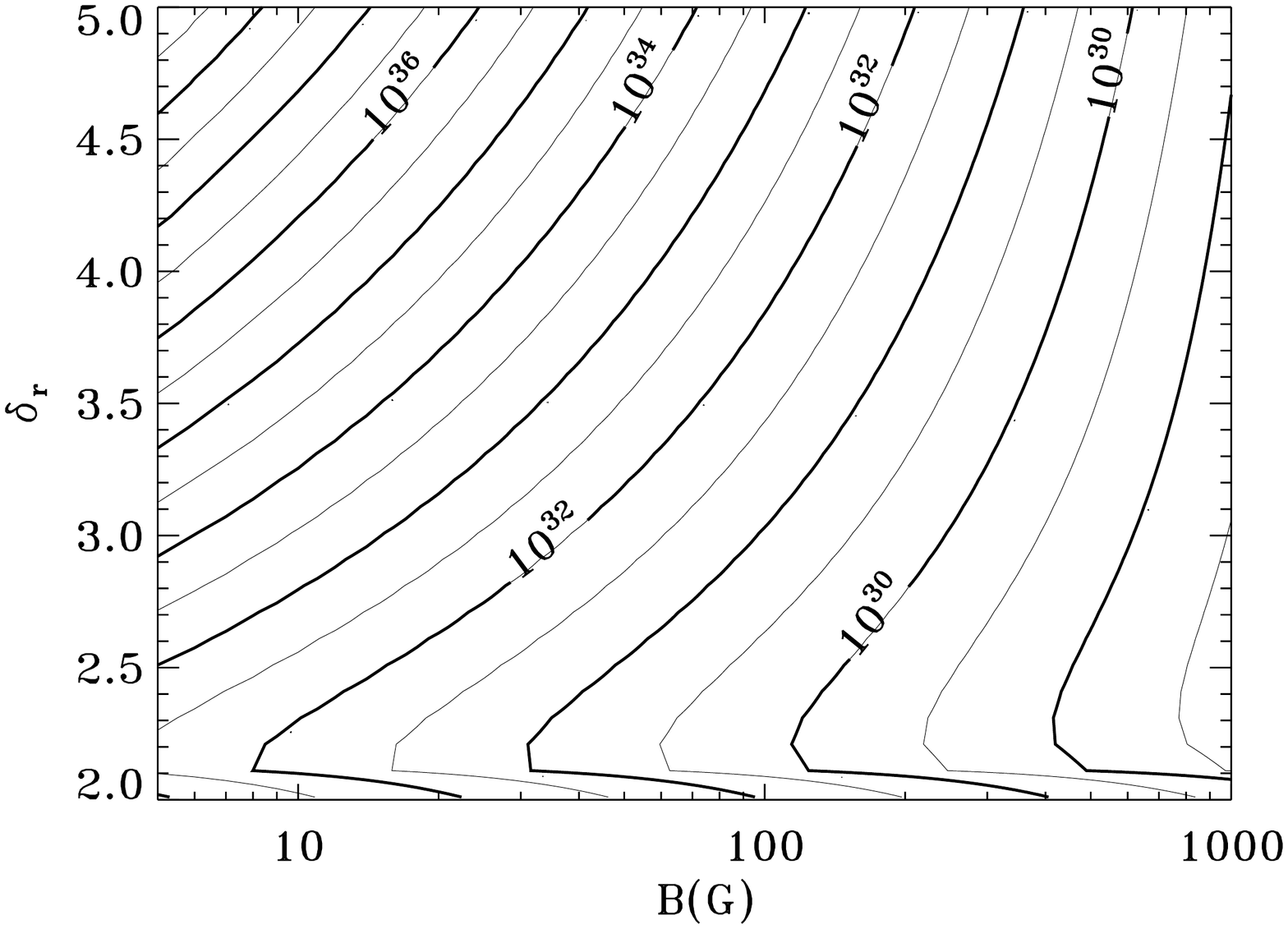}
\vspace*{-1.cm}
\caption{
(left) Modelled light curve of a short duration (lasting 5 minutes), small enhancement  (factor of three increase
above quiescence) flare on a nearby (5 pc) star. Right y-axis gives intrinsic radio luminosity, left axis
gives the estimated radio flux density at 17 GHz. Noise is estimated in each 1 minute bin by offsetting the
flux by a random number times the expected rms values.
(right)
Contour plot of $\delta_{r}$, the index of accelerated particles, and $B$, the magnetic field strength in the
radio-emitting source, for the radio flare seen at left, for different kinetic energies in the flare. The ngVLA would be able to constrain the distribution of the accelerated
particles as well as the magnetic field strength in the radio-emitting source through equipartition with radiated flare energies.
\label{fig:ptcls}}
\end{figure}

 Such an example is shown in the left panel of Figure~\ref{fig:ptcls}.
 These types of observations would be most sensitive to the nearest M dwarfs, in order to study time-variable emission
 over very short timescales (minutes). Quiescent radio luminosities as low as 10$^{11}$ erg s$^{-1}$ Hz$^{-1}$ (at the limit of current detection capabilities; Osten 
 et al. 2015) would be detectable in short integrations, allowing for study of small enhancement events only factors of a few larger.
 The kinetic energies probed here are moderate to large sized solar flares, and probe magnetic field strengths
 in the tens to hundreds of Gauss level.


  \section{Summary}
 
 The influence that stars have on their near environments is a timely research topic now, and is expected
 only to grow in the future as exoplanet discoveries grow, and exoplanet science moves beyond detection and characterization of individual objects, to a fuller consideration of the impact of the planetary environment.  While we expect that new discoveries 
 in this area in the next decade will fill in some areas where we currently lack insight, there will remain gaps in our knowledge which can only be filled by a next generation sensitive radio telescope operating in the deka-GHz range.
 
 \section{Acknowledgements}
 RAO and MKC acknowledge support from an NRAO Community Studies Report for the research reported here. 
JL acknowledges support from the Jet Propulsion Laboratory, California Institute of Technology, under a contract with the National Aeronautics and Space Administration. 

\section{References}
\begin{multicols}{2}
\noindent 
    Fichtinger, B. et al. 2017 A\&A 599, 127\\
Heyner, D. et al. 2012 ApJ 750, 133 \\
  Jackman, C. H. et al. 1990 JGR 95, 7417 \\ 
  Johnstone, C. P. et al. 2015 A\&A 577, 28\\
  Lovelace, R. et al. 2008 MNRAS 389, 1233 \\
  Osten, R. A. et al. 2015 ApJ 805, L3 \\
  Osten, R. A. et al. 2016 ApJ 832, 174\\
  Osten, R. A. \& Wolk, S. J. 2017 IAUS 328, 243\\
  Segura, A. et al. 2010 AsBio 10, 751\\
  Smith, K. et al. 2005 A\&A 436, 241\\
  Tu, L. et al. 2015 A\&A 577, 3 \\
  Wargelin, B. J. \& Drake, J. 2001 ApJ 546, L57\\
  Wood, B. E. et al. 2014 ApJ 781, L33\\
  \end{multicols}

\end{document}